\documentclass[twocolumn,twoside,showpacs,showkeys,
  preprintnumbers,superscriptaddress,amsmath,amssymb,pra]{revtex4}
\usepackage[dvips]{graphicx}
\usepackage[dvips]{color}

\newcommand{\definedas}{\ensuremath{\stackrel{\mathrm{def}}{=}}}
\newcommand{\id}{\ensuremath{{\mathbb{I}}}}
\newcommand{\reni}{R\'enyi }
\newcommand{\ket}[1]{\ensuremath{|#1\rangle}}
\newcommand{\bra}[1]{\ensuremath{\langle#1|}}
\newcommand{\braket}[2]{\ensuremath{\langle#1|#2\rangle}}
\newcommand{\ketbra}[2]{\ensuremath{|#1\rangle\langle#2|}}
\newcommand{\uo}{\ket{u}}
\newcommand{\ui}{\ket{\bar{u}}}
\newcommand{\vo}{\ket{v}}
\newcommand{\vi}{\ket{\bar{v}}}
\newcommand{\ww}{\ket{w}}
\newcommand{\e}[1]{\ket{e_{#1}}}
\newcommand{\eo}{\e{0}}
\newcommand{\ei}{\e{1}}

\newcommand{\kpsi}[1]{\ket{\psi_{#1}}}
\newcommand{\bkpsi}[2]{\braket{\psi_{#1}}{\psi_{#2}}}
\newcommand{\npsi}[1]{\ensuremath{\| \psi_{#1} \|}}
\newcommand{\bPHI}[1]{\bra{\Phi_{#1}}}
\newcommand{\kPHI}[1]{\ket{\Phi_{#1}}}
\newcommand{\bkPHI}[2]{\braket{\Phi_{#1}}{\Phi_{#2}}}
\newcommand{\nPHI}[1]{\ensuremath{\| \Phi_{#1} \|}}
\newcommand{\w}[1]{\ket{w_{#1}}}
\newcommand{\cpi}{\ensuremath{\langle P_c^1 \rangle}}
\newcommand{\cpiE}{\ensuremath{\langle P_{c=}^1 \rangle}}
\newcommand{\cpiD}{\ensuremath{\langle P_{c\neq}^1 \rangle}}
\newcommand{\perr}{\ensuremath{P_{\mathrm{err}}}}
\renewcommand{\log}{\ensuremath{\mathrm{log}_2}}
\newcommand{\feq}{\ensuremath{f_{[=]}}}
\newcommand{\fneq}{\ensuremath{f_{[\neq]}}}
\newcommand{\tr}{\operatorname{Tr}\nolimits}
\newcommand{\sigcond}[1]{\ensuremath{\sigma_{#1}}}
\newcommand{\takethisout}[1]{}

\begin{document} 

\title{On the optimality of individual entangling-probe attacks\\
  against BB84 quantum key distribution}

\newcommand{\Faculty}{Quantum Optics, Quantum Nanophysics and
  Quantum Information, Faculty of Physics, University of Vienna,
  Boltzmanngasse~5, 1090 Vienna, Austria}
\newcommand{\ARC}{Austrian Research Centers~GmbH~-~ARC,
  Donau-City-Str.~1, 1220 Vienna, Austria}

\author{Isabelle Herbauts}
\email[Corresponding author:~]{isabelle.herbauts@univie.ac.at}
\affiliation{\Faculty}
\author{Stefano Bettelli}
\affiliation{\ARC}
\author{Hannes H\"{u}bel}
\affiliation{\Faculty}
\author{Momtchil Peev}
\affiliation{\ARC}

\preprint{arXiv:0711.1728v1}

\date{October the 19th, 2007}
\begin{abstract}
  
Some MIT researchers \cite{KSWS07} have recently claimed that their
implementation of the Slutsky-Brandt attack \cite{SRSF98, Brandt05} to the
BB84 quantum-key-distribution (QKD) protocol puts the security of this
protocol {\em ``to the test''} by simulating {\em ``the most powerful
individual-photon attack''} \cite{ShapiroWong06}. \takethisout{However, the
interpretation of the consequences of this attack by the authors is
somewhat misleading; moreover, } A related unfortunate news feature by a
scientific journal \cite{Brumfiel_NewsNature07, Brumfiel_FeatureNature07}
has spurred some concern in the QKD community and among the general public
by misinterpreting the implications of this work. The present article
proves the existence of a stronger individual attack on QKD protocols with
encrypted error correction, for which tight bounds are shown, and clarifies
why the claims of the news feature incorrectly suggest a contradiction with
the established ``old-style'' theory of BB84 individual attacks.

The full implementation of a quantum cryptographic protocol includes a
reconciliation and a privacy-amplification stage, whose choice alters in
general both the maximum extractable secret and the optimal eavesdropping
attack. The authors of \cite{KSWS07} are concerned only with the error-free
part of the so-called sifted string, and do not consider faulty bits,
which, in the version of their protocol, are discarded. When using the
provably superior reconciliation approach of encrypted error correction
(instead of error discard), the Slutsky-Brandt attack is no more optimal
and does not ``threaten'' the security bound derived by L\"utkenhaus
\cite{Lutkenhaus99}.

It is shown that the method of Slutsky and collaborators \cite{SRSF98} can
be adapted to reconciliation with error correction, and that the optimal
entangling probe can be explicitly found. Moreover, this attack fills
L\"utkenhaus bound, proving that it is tight (a fact which was not
previously known).

\end{abstract}

\maketitle

\section{Introduction}

Quantum cryptography, or, more properly, quantum key distribution (QKD) is
a discipline investigating techniques to grow, out of a common secret key,
a larger key between two remote parties (Alice and Bob) linked by a quantum
and a classical communication channel. The generated key can then be
consumed to perform various classical cryptographic tasks, such as encoding
messages with a one-time pad, but this is outside the scope of QKD. In the
last twenty years it has been shown that it is in principle possible to
grow the secret despite the channels being under the control of a
non-disruptive attacker (Eve) subject only to the laws of quantum
mechanics, a task deemed impossible in a completely classical setting; this
ability stems ultimately from the well-known tradeoff between acquired
knowledge and state disturbance in a quantum measurement. For an
introduction to the subject, the interested reader is pointed to some
recent \cite{GRTZ02, DLH06} and forthcoming \cite{SBPCDL08} reviews.

Broadly speaking, QKD protocols are based on Alice transmitting quantum
systems (usually photons) in randomly selected states out of an alphabet of
nonorthogonal states. When Bob receives a system, he performs a measurement
to infer Alice's signal; at the end of the quantum exchange, the
measurement settings (but not the results) are publicly compared, and only
results from compatible measurements are retained (key sifting). In the
sifted key, measurement results are ideally deterministically correlated,
and any eavesdropping activity, which fundamentally disturbs the exchanged
systems, can be monitored.  The oldest and best studied QKD procedure,
described later on, is known under the name of Bennett-Brassard 1984 (BB84)
protocol \cite{BennettBrassard84}; other procedures, very similar in spirit
to BB84, are the entanglement-based Ekert \cite{Ekert91} and BBM92
\cite{BBM92} protocols.

QKD protocols so far devised consist of {\small (a)} a quantum transmission
followed by sifting over a public authenticated classical channel,
establishing a highly correlated pair of keys at two remote sites; {\small
(b)} a reconciliation procedure over the classical channel, allowing Alice
and Bob to agree on a shared identical random key; {\small (c)} a
privacy-amplification procedure over the classical channel which ensures
the security of a shortened key obtained from the sifted key
\cite{Maurer93, BBCM95}. An additional necessary task for a complete secure
protocol is authentication, but this is of no major consequence in the
present analysis. Since the bits of the raw key are all statistically
independent, no information about the sifted key can be extracted from the
discarded bits of the raw key, and therefore general security analyses are
concerned only with sifted keys. In both the reconciliation and the privacy
amplification phases, however, information is exchanged over the classical
authenticated channel, which can be perfectly spied, although not modified,
by Eve. This is to be taken into account, in order that, after a sequence
of appropriate procedures, both Alice and Bob possess a copy of a key, about
which Eve knows only a negligible amount of information. The security of a
QKD protocol, therefore, relates directly to a quantitative estimation of
the amount of information potentially acquired by Eve on the sifted and
reconciled key.

The conditions for the security of full QKD protocols have been extensively
studied; in general, they depend on the class of allowed attacks and on the
degree of non-ideality of the involved channels and cryptographic devices.
In this article only individual attacks, where Eve is restricted to
interact with and measure each transmitted signal independently, are
considered; moreover, the channel is assumed to be noisy and potentially
leaking, but the other devices are ideal and the quantum exchange is
analysed only in the limit of very large keys. In this scenario, security
conditions are often expressed in the form of a discarded fraction
$\tau(e)$, that is the portion of the sifted and reconciled key that is to
be sacrificed in order to obtain a final secret key. The discarded fraction
is a function of the probability that a bit at Alice's site and the
corresponding bit at Bob's site differ after sifting, {\em i.e.}, the
quantum-bit error rate (QBER) $e$; in the usual conservative approach, it
must be assumed that errors in the sifted key are entirely due to Eve.

Admittedly, this is not the state of the art in QKD security proofs, since
the most general class, where all signals are made to interact coherently
with a very large probe which is then optimally measured by Eve (coherent
attacks), has already been tackled \cite{ILM01, GLLP04}. Also, scenarios
where Alice and Bob's devices are imperfect and potentially manipulated by
Eve have been considered and partially analysed, as well as the case of
finite lengths for the exchanged keys. Finally, in recent years the
definition itself of what is a secure final key has changed, due to the
introduction of the notion of composability. Literature on these subjects
is too large to be even cited here; the interested reader should refer to
\cite{SBPCDL08}. 

It must be remarked, however, that the case of ideal individual attacks
still bears some importance because {\small (a)} proofs for realistic
devices and finite key lengths are ultimately based on proofs for ideal
ones; {\small (b)} security bounds for individual attacks, although
conceptually very different, give results rather similar to the case of
coherent attacks, which is a convincing argument about the effectiveness of
eavesdropping strategies for those researchers that see coherent attacks as
technologically unfeasible; and {\small (c)} individual attacks are a
sufficiently simple class to be readily understood by researchers working
on practical implementations, and their complete understanding helps
dissipating that aura of phenomenologicality which is sometimes associated
to security bounds in actual QKD protocols (as if a security bound, which
is a purely mathematical statement and not an observable, could be subject
to experimental investigation).

Recently, Kim \textit{et al.} \cite{KSWS07} have claimed to physically
implement {\em ``the most powerful individual-photon attack''}, therefore
putting the BB84 protocol's security {\em ``to the test''}
\cite{ShapiroWong06}. Following their suggestion that {\em ``the physical
simulation allows investigation of the fundamental security limit of the
BB84 protocol against eavesdropping in the presence of realistic physical
errors, and it affords the opportunity to study the effectiveness of error
correction and privacy amplification when the BB84 protocol is attacked''},
in this article this particular attack \cite{SRSF98, Brandt05} (from now
on, the {\em Slutsky-Brandt attack}, (SB)\footnote{It is to be remarked
that, despite the name of Slutsky being used for this attack in literature,
the authors of \cite{SRSF98} have never overclaimed its optimality.}) is
analysed in the context of a complete and efficient QKD protocol.

For individual eavesdropping attacks, and using an appropriate
reconciliation protocol which does not correlate signals, upper bounds on
Eve's information can be estimated via the average collision probability of
the sifted key. A security bound as a function of the disturbance has been
derived by L\"utkenhaus \cite{Lutkenhaus99} in both scenarios when faulty
bits are discarded or corrected, by modelling Eve's individual attacks by
means of positive-operator-valued measurements (POVM). In sections
\ref{sec:Slutsky_approach} and \ref{sec:Slutsky_approach_with_ED} the SB
attack is analysed, and it is highlighted that this attack yields the upper
value of $\tau(e)$, the discarded fraction in the privacy amplification
stage, obtained by L\"utkenhaus when faulty bits are rejected, therefore
conferring L\"utkenhaus bound the property of being sharp, as already
pointed out by this author.

However, the BB84 dialect that is nowadays most commonly adopted implements
the reconciliation step through error correction (instead of error
discard), because this leads to a larger final secret key, as shown in
section \ref{sec:reconciliation}. During this procedure, assumed perfect
for simplicity, an amount $h(e)$ of information per sifted bit (the Shannon
limit \cite{Shannon48}) is leaked to Eve and must be discarded. In section
\ref{sec:Slutsky_Brandt_with_EC}, it is shown that for such protocol the SB
attack is not necessarily optimal, and in no way threatens the upper bound
on $\tau(e)$ as derived by L\"utkenhaus for individual attacks on QKD
protocols with error correction \cite{Lutkenhaus99}.

Finally, in section \ref{sec:optimal_with_EC_leakage}, it is proven that
there exist a stronger entangling-probe attack, and that this attack leads
to a discarded fraction that coincides exactly with L\"utkenhaus upper
bound, thus abrogating the regime of hope for individual attacks against an
ideal BB84 protocol with encrypted error correction subsisting thus far.

The mathematical techniques used in this article are similar to those
developed by \cite{FuchsPeres96, FGGNP97} and perfected in \cite{SRSF98},
but an important extension is introduced in Sec.(\ref{sec:parametrisation})
which allows for a significant simplification of the problem. The final
result suggests an intriguing relation between the maximal collision
probability achievable through an optimal measurement and the fidelity of
the (mixed) states to be distinguished.

\section{Modelling of individual attacks and security bounds}
\label{sec:Slutsky_approach}

In general, a security proof for a given class of attacks is made out of
three main ingredients. First, one needs a mathematical description (a
parametrisation) of all elements of the class. Then, one must estimate how
dangerous each element is with respect to the final goal of establishing a
secret key shared by Alice and Bob; this very much relies on the definition
of security, and usually takes the form of non-tight bounds. Last, an
optimisation is to be performed in the parametrised attack space in order
to bound the power of the most threatening element for each value of the
disturbance parameter ({\em e.g.}, the QBER). The first two steps in the
case of ideal individual attacks, according to the approach of Slutsky,
Rao, Sun and Fainman \cite{SRSF98}, are reviewed in this section.

\subsection{The entangling-probe model}
\label{sec:model}

In 1996 Fuchs and Peres \cite{FuchsPeres96,FGGNP97} introduced the
following individual-attack model. Eve prepares a probe and lets it
interact with the signal system sent by Alice; the joint unitary evolution
leaves the two systems in an entangled quantum state. The signal is then
forwarded to Bob, while the probe is stored by Eve and measured after the
reconciliation stage. Entanglement between the system and the probe
``induces'' a correlation between Eve's and Bob's measurements, allowing
Eve to obtain partial information on the key. This model is known as
Fuchs-Peres' entangling-probe (FPEP) attack.

The definition of individual attack does not prevent Eve from forwarding to
Bob a system with a different Hilbert space from the original one, a case
not covered by the FPEP model.\footnote{For instance, if the signal system
is a photon and the channel is an optical fibre, Eve could inject
additional photons in the fibre to fool Bob's detectors.} It has however
been shown \cite{LutkenhausPhD96, Lutkenhaus99, WZY02} that, if Bob's
apparatus can, to some extent, reveal the presence of multiple systems in
the signal, by adding a sufficiently large penalty to the QBER in case of
multiple detections it is always possible to render these attacks
non-optimal for Eve.\footnote{In practice, it is sufficient to insert a
random bit in the sifted string for each multiple detection instead of
neglecting that detection.}

That the FPEP model indeed covers the full class of individual attacks (at
least among attacks where Eve is forced to measure its system at some
point) is a consequence of Stinespring's dilation theorem
\cite{Stinespring55}, that guarantees that every completely positive and
trace-preserving map can be built by embedding the input state space in the
state space of a ``larger'' system, which is then unitarily evolved and
subsequently traced down to a subsystem isomorphic to the output
space. Therefore, any quantum channel can be regarded as arising from a
unitary evolution on a larger (dilated) system. Embedding in a larger space
can be thought of as tensoring with an auxiliary system (the probe) in a
fixed initial state, because this provides an intuitive physical model. The
initial state can moreover be assumed to be pure.\footnote{The ability to
purify each mixed state into a pure state of a larger system is again a
consequence of Stinespring's theorem.} Stinespring's theorem is a
generalisation of Neumark's theorem \cite{Neumark40}, that shows that every
generalised measurement on a system can be implemented by letting the
system interact unitarily with an ancilla, and then projectively measuring
the latter.\footnote{In reality, the theorem asserts that a general
measurements with $n'$ possible outcomes on an $n$-dimensional system, with
$n'>n$, can always be seen as a projective measurement on an enlarged space
with $n'$ dimensions which embeds the original state space. But the version
with the measurement only on the auxiliary system is easier to visualise.}

The explicit FPEP parametrisation for the BB84 protocol will now be
introduced, following the notation of \cite{SRSF98} as closely as possible.
In BB84, Alice randomly chooses a basis from a pair $\{ \uo, \ui \}$ and
$\{ \vo, \vi \}$ of mutually unbiased orthogonal bases, and a signal bit,
and sends to Bob the first element of the basis if the chosen bit is $0$,
the second element otherwise. Bob, similarly, chooses, randomly and
independently from Alice, one of the two bases, and performs a von Neumann
measurement to determine the bit. The sifted key is built from those
exchanges where the measurements were compatible, {\em i.e.}, when both
Alice and Bob chose the same basis.

If $U$ is the unitary joint evolution of the FPEP attack, and $\ww$ is the
initial pure state of the probe, the overall entangled state after
interaction can be decomposed as
\begin{equation} \label{eq:joint_evolution}
  U \ket{a} \ww = \ket{a} \kpsi{aa} + \ket{\bar{a}} \kpsi{a\bar{a}},
\end{equation}
where $a \in \{ u, \bar{u}, v, \bar{v} \}$, and $\ket{\bar{a}}$ is the
state corresponding to the complementary bit (the states $\kpsi{ab}$ are
neither orthogonal nor normalised). When the input state $\ket{a}$ is sent
by Alice, every outcome $b$ of Bob is therefore associated to an output
state of the probe proportional to $\kpsi{ab}$. It is convenient
\cite{FuchsPeres96, SRSF98} to define an orthonormal basis $\{ \eo, \ei
\}$, oriented symmetrically with respect to the signal states, which can
then be expressed as
\begin{subequations} \label{eq:def_of_states}
  \newcommand{\COS}{\makebox[6ex]{$\cos\alpha$}} 
  \newcommand{\SIN}{\makebox[6ex]{$\sin\alpha$}}
  \begin{align} 
    \uo &= + \COS \eo + \SIN \ei, \\
    \ui &= - \SIN \eo + \COS \ei, \\
    \vo &= + \SIN \eo + \COS \ei, \\
    \vi &= + \COS \eo - \SIN \ei,
  \end{align}
\end{subequations}
where $\alpha = \pi/8$, because the bases are unbiased. Since $\eo$ and
$\ei$ generate the signal space, the action of a generic FPEP attack is
then fully described by the action of $U$ on them; similarly to
Eq.\ref{eq:joint_evolution}, one defines
\begin{equation} \label{eq:joint_evolution_e}
  U \e{m} \ww = \eo \kPHI{m0} + \ei \kPHI{m1}.
\end{equation}
As for the $\kpsi{ab}$'s, the four states $\kPHI{mn}$ are generally neither
normalised nor orthogonal; their number shows that the probe space
corresponding to a two-level signal is effectively four-dimensional.

\subsection{Attack-space refinement via symmetrisation}
\label{sec:symmetrisation}

The aforementioned space of attacks is by far too complicated to be
completely explored. However, standard techniques based on symmetrisation
are available to reduce its size without loosing potential optimal
elements. The general idea is trivial: if a subset of the space is known
where all attacks are {\em equivalent}, it is sufficient to retain only one
representant of the subset during the search. What is less trivial is how
to characterise and find equivalent elements. In the picture of the
entangling-probe, all measurable quantities are determined by the joint
state $\chi$ of the signal and probe after interaction. If $\rho_a \in \{
\rho_u, \rho_{\bar{u}}, \rho_v, \rho_{\bar{v}} \}$ is a signal state and
$\omega = \ket{w}\bra{w}$ is the initial probe state, then
\begin{equation}
  \chi(\rho_a, \omega, U) = U \rho_a\otimes\omega\, U^\dagger,
\end{equation}
The effects of an attack $(U, \omega)$, both in terms of the QBER and Eve's
maximum inference power, are summarised by the statistical distribution of
the $\chi$'s, which depends on the signal a-priori distribution $p_a$, that is
\begin{equation}
  (U, \omega) \longleftrightarrow \{\, p_a;\, \chi(\rho_a, \omega, U) 
  \,\}_{a = u, \bar{u}, v, \bar{v}}.
\end{equation}

Since, for BB84, the a-priori probabilities $p_a = 1/4$ are the same,
attacks to the protocol have equivalent effects if the rays of the states
are permuted (without violating the constraint that the two bases are
unbiased). Readers not interested in technicalities may now just retain
that the simplification of the search space implies that the vectors
$\kpsi{}$ of Eq.(\ref{eq:joint_evolution}) can be parametrised with only
two real parameters, and jump to Eqs.(\ref{eq:values_of_bkpsi_simple}) in
Sec.(\ref{sec:parametrisation}).

All ray permutations can be generated with only two involutions, for
instance {\small (1)} the basis exchange and {\small (2)} the bit exchange
in the second basis; these two specific symmetries are called in the
following respectively $R_1$ and $R_2$. However, the approach is more
general, and can be extended to other cases, for example to the six-state
variant of BB84 \cite{Bruss98}.

Let $Q_i = R_i \otimes \,\id$ be a local operator on the joint space of the
signal and the probe\footnote{A local operator can be implemented without
communication by Eve and Bob in their laboratories. Note that one could
also define $Q_i = R_i \otimes R_i^{\prime}$ with a generic unitary
transformation $R_i^{\prime}$ in Eve's space, since every such
transformation could be undone by Eve during her optimal measurement; but
this degree of freedom does not bring additional constraints and is thus
ignored here.}; if Alice changes her signal convention from $\rho_a$ into
$R_i \rho_a R_i^\dagger$, and the final density matrix $\chi(R_i \rho_a
R_i^\dagger, \omega, U)$ is transformed back in Bob's laboratory into
$Q_i^\dagger \chi \,Q_i$, both the QBER and Eve's maximum inference power,
which are average quantities, are statistically unchanged. It follows, very
much in analogy to the passive-active picture of a reference-frame change,
that the attacks $(U, \omega)$ and ($Q_i^\dagger U Q_i, \omega)$ are
equivalent. In mathematical terms
\begin{multline} \label{eq:equivalent_U}
  \chi(\rho, \omega, U) \equiv 
  Q_i^\dagger \chi(R_i \rho R_i^\dagger, \omega, U)\, Q_i \\
  = (Q_i^\dagger U Q_i) \rho\otimes\omega\, (Q_i^\dagger U Q_i)^\dagger
  = \chi(\rho, \omega, Q_i^\dagger U Q_i).
\end{multline}

\begin{figure}
  \begin{center}
    \includegraphics[width=.65\linewidth]{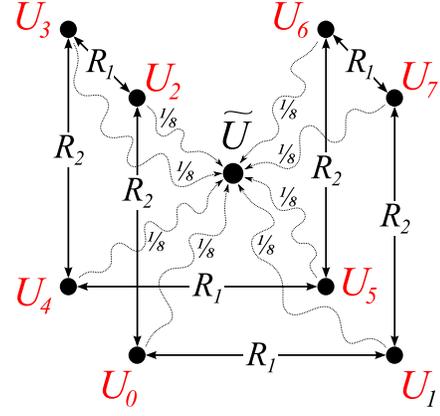}
  \end{center}
  \caption{\label{fig:symmetries} A graphical representation of the orbit
    $U_{g\in[1\dots 8]}$ generated by applying the symmetry group of the
    BB84 protocol to a generic attack $U$. The whole orbit can be explored
    using only the involutions $R_1$ (basis-exchange) and $R_2$
    (bit-exchange). The attack $\widetilde U$ is the average of the
    elements on the orbit, operates on an enlarged probe space and is
    symmetric under the BB84 group. The search for optimal elements can be
    restricted to these symmetric attacks.}
\end{figure}

Therefore, there is a direct link between a representation of the group $G$
of symmetries of the protocol and attack equivalence, and this remark can
be exploited in a useful way. Below we consider the case of finite $G$,
which is proper to the BB84 protocol. Since $R_1$ and $R_2$ generate the
whole representation, by repeated application of Eq.(\ref{eq:equivalent_U})
it can be shown that, for all $R_g$, the attack $U_g = Q_g^\dagger U Q_g$
is equivalent to $U = U_0$ ($\omega$ is omitted here, since it is always
the same, and $Q_{g\in G} = R_g \otimes \id$). For BB84, the relevant group
$G$ is $D_4$ \cite[chap. XII, table 7]{LandauLifshitzIII}; the action of
the representation is illustrated in Fig.(\ref {fig:symmetries}). The order
of the group is $8$, so that the orbit of $U$ has {\em at most} $8$
elements.

Intuitively, a random application by Eve of attacks $U_g$ will give another
equivalent attack. The idea can be formalised by extending the probe space
with an auxiliary space with $|G|$ dimensions. Define
\begin{equation}
  \widetilde U = {\textstyle\sum_g}\,\, U_g \otimes P_g
  \qquad\textrm{and}\qquad \widetilde\omega = \omega \otimes \Omega,
\end{equation}
where $P_g = \ket{g}\bra{g}$ are orthogonal projectors in the auxiliary
space, and $\Omega = |G|^{-1} \sum_{gg'} \ket{g}\bra{g'}$ is the density
matrix of a pure state with $\tr(P_g \Omega) = 1/|G|$. The connection with
the intuitive idea is that the projectors in the auxiliary space randomly
select the $U_g$'s; the construction of $(\widetilde U, \widetilde \omega)$
is represented in Fig.(\ref{fig:symmetries}).

What is special about the ``average'' attack $(\widetilde U, \widetilde
\omega)$ built in this way is that it is invariant under a group, which can
be built from the representation of $G$ and some permutation operators
$X_g$ on the auxiliary space. Let
\begin{equation} \label{eq:def_tildeR}
  \widetilde R_g \definedas Q_g^\dagger \otimes X_g 
  = R_g^\dagger \otimes \id \otimes X_g
  \definedas R_g^\dagger \otimes \widehat{R}_g.
\end{equation}
Operators $X_g$ are chosen such that if $Q_g^\dagger U_{\ell} Q_g \!=
U_{\pi_g(\ell)}$ then $X_g P_{\ell} X_g^\dagger = \!P_{\pi_g(\ell)}$. This
is always possible due to the fundamental theorem
\cite[chap. XII]{LandauLifshitzIII} that any finite group of order $k$ is
isomorphic to a subgroup of the general symmetric group of all permutations
of $k$ elements, $S(k)$, which in turn can be naturally represented by the
set of all $k \times k$ permutation matrices. It is then sufficient to fix
one isomorphism and chose $X_g$ as the isomorphic image of $Q_g^\dagger$;
in this way $X_g \ket{\ell} = \ket{\pi_g(\ell)}$. It is now a trivial
matter to verify that
\begin{align}
  \widetilde R_g \widetilde U {\widetilde R_g}^\dagger 
  &= {\textstyle\sum_{\ell}} \, Q_g^\dagger U_{\ell} Q_g
  \otimes X_g P_{\ell} X_g^\dagger = \widetilde U \\
  \textrm{and} \quad \widehat{R}_g \widetilde \omega \widehat{R}_g^\dagger
  &= \omega \otimes X_g \Omega X_g^\dagger = \widetilde \omega.
\end{align}
One can therefore conclude that, given a group $G$ of protocol symmetries,
for each attack $(U, \omega)$ there exists an equivalent attack
$(\widetilde U, \widetilde \omega)$ which is invariant under all
$\widetilde R_g$'s as defined in Eq.(\ref{eq:def_tildeR}). It follows that
the subset containing all attacks invariant under such symmetries contains
at least one optimal element; the search for optimality can thus be
restricted to that subset. This finding is directly relevant to the FPEP
parametrisation, because it generates constraints for the $\kPHI{mn}$'s of
Eq.(\ref{eq:joint_evolution_e}). In fact, for invariant attacks, replacing
$U$ with $(R_g^\dagger \otimes \widehat R_g) \, U (R_g^\dagger \otimes
\widehat R_g)^\dagger$ and $\ket{\omega}$ with $\widehat R_g \ket{\omega}$
shows that
\begin{equation} \label{eq:action_of_Rp}
  U R_g \e{m} \ket{\omega} 
  = \sum_n R_g \e{n} \widehat R_g^\dagger \kPHI{mn},
\end{equation}
from which, for each symmetry $R_g$, the value of $\widehat R_g^\dagger
\kPHI{mn}$ can be calculated and used in constraints of the form
\begin{equation} \label{eq:constraint_PHI}
  \bkPHI{mn}{pq} = \bPHI{mn} \widehat R_g \widehat R_g^\dagger \kPHI{pq}.
\end{equation}
This formula is clearly valid for all $g\in G$, but in practice it is
sufficient to restrict its application to $\widehat R_1$ and $\widehat
R_2$. Also, it is more convenient to work with the symmetries of the state
vectors $\ket{a}$ instead of those of the corresponding rays. This gives a
representation of $D_8$ instead of $D_4$, where redundant elements are
included (like $\ket{a} \rightarrow -\ket{a}$, which is physically
indistinguishable from the identity); the generated constraints are however
the same.

\subsection{The entangling-probe parametrisation}
\label{sec:parametrisation}

The authors of the FPEP model remarked that the BB84 protocol, as described
above, is endowed with the basis-exchange symmetries $R_1$ (an involution
corresponding to $\eo \leftrightarrow \ei$). Then, using essentially the
same techniques described in Sec.(\ref{sec:symmetrisation}), namely
Eq.(\ref{eq:constraint_PHI}), they showed that an attack-dependent
orthonormal basis $\{ \w{i} \}_{i \in 0 \dots 3}$ can be found\footnote{The
absence of coefficient $X_4$ is due to backward compatibility.} such that
\begin{subequations} \label{eq:def_of_Phi}
  \newcommand{\fterm}[1]{\makebox[10ex]{$#1$}}
  \begin{align}
    \kPHI{00} &= \fterm{X_0 \w{0}\,+}  X_1 \w{1} + X_2 \w{2} + X_3 \w{3}, \\
    \kPHI{01} &= \fterm{            }  X_5 \w{1} + X_6 \w{2}, \\
    \kPHI{10} &= \fterm{            }  X_6 \w{1} + X_5 \w{2}, \\
    \kPHI{11} &= \fterm{X_3 \w{0}\,+}  X_2 \w{1} + X_1 \w{2} + X_0 \w{3}.
  \end{align}
\end{subequations}
With analogous considerations extended to anti-unitary symmetries (complex
conjugation in the probe space) they also showed that all coefficients $X$
are real numbers. Note that this parametrisation satisfies $\bkPHI{mn}{pq}
= \bkPHI{\bar m \bar n}{\bar p \bar q} = \bkPHI{pq}{mn}$, given by the
constraints of $\smash{\widehat R_1}$ (as previously, the bar indicates the
complementary bit). The $X$'s are correlated by the fact that $U$ must be a
unitary operator, hence the additional constraints
\begin{subequations}
  \begin{align}
    1 &= {\textstyle\sum_{i=0,1,2,3,5,6}} \,X_i^2 
    = \nPHI{00}^2 + \nPHI{01}^2, \\
    0 &= X_1 X_6 + X_2 X_5 = \bkPHI{00}{10} = \bkPHI{11}{01};
    \label{eq:orthogonality_PHI0010}
  \end{align}
\end{subequations}
this shows that each FPEP attack, prior to Eve's measurement, can be
described by only four real parameters.

However, as already said, there exists another symmetry in the BB84
protocol which has not been exploited by the authors of \cite{SRSF98},
namely $R_2$, the bit-exchange symmetry in one basis only. This corresponds
to swapping the convention for $0$ and $1$ in one basis while leaving the
other convention unchanged. The bit-exchange symmetry is generated by a
Hadamard transformation:
\begin{equation}
  \left[ \begin{aligned} \eo \\ \ei \end{aligned} \right] \longrightarrow
  \begin{pmatrix} 1 & \hphantom{-}1 \\ 1 & -1 \end{pmatrix}
  \left[ \begin{aligned} \eo \\ \ei \end{aligned} \right].
\end{equation}
It is easy to check that $\vo \leftrightarrow \vi$, while $\uo$ and $\ui$
are invariant (actually, $\ui$ has its sign flipped, but this does not
matter, since the physical state is the same). Using $R_2 \e{j} = [ \eo +
(-1)^j \ei ] / \sqrt{2}$, after some elementary algebraic passages, using
Eq.(\ref{eq:action_of_Rp}), one obtains
\begin{subequations} \label{eq:output_RpPhi}
  \begin{align}
    \widehat R_2 \kPHI{00} &= \tfrac{1}{2} 
    \left( \kPHI{00} + \kPHI{01} + \kPHI{10} + \kPHI{11} \right), \\
    \widehat R_2 \kPHI{01} &= \tfrac{1}{2} 
    \left( \kPHI{00} - \kPHI{01} + \kPHI{10} - \kPHI{11} \right), \\
    \widehat R_2 \kPHI{10} &= \tfrac{1}{2} 
    \left( \kPHI{00} + \kPHI{01} - \kPHI{10} - \kPHI{11} \right), \\
    \widehat R_2 \kPHI{11} &= \tfrac{1}{2} 
    \left( \kPHI{00} - \kPHI{01} - \kPHI{10} + \kPHI{11} \right).
  \end{align}
\end{subequations}

Eq.(\ref{eq:constraint_PHI}) shows how to use these relations to calculate
additional constraints for $\bkPHI{mn}{pq}$ products. Of course, not all
combinations of indexes are interesting, because quite a few are already
fixed by other symmetries and the unitarity of $U$. As already seen, there
are at most four ``independent'' products, {\em e.g.}, $\bkPHI{00}{01}$,
$\bkPHI{01}{01}$, $\bkPHI{00}{11}$, and $\bkPHI{01}{10}$. The most
important constraint is obtained by calculating the first one,
\begin{equation} \label{eq:c_is_0}
  \bkPHI{00}{01} = \bPHI{00} \widehat R_2 \widehat R_2^\dagger \kPHI{01} 
  = X_1 X_5 + X_2 X_6 = 0.
\end{equation}

Together with Eq.(\ref{eq:orthogonality_PHI0010}), this relation proves a
fundamental property of the probe space for optimal attacks, {\em i.e.},
this space is the direct sum of two orthogonal subspaces, one corresponding
to bits received correctly by Bob and the other to errors in the sifted key,
\begin{equation}
  \textrm{Span} \left\{ \kPHI{00}, \kPHI{11} \right\} \perp 
  \textrm{Span} \left\{ \kPHI{01}, \kPHI{10} \right\}.
\end{equation}
The symmetries analysed so far have also led to the conclusion that, within
each subspace, basis vectors have the same length, $\nPHI{00} \!=\!
\nPHI{11}$ and $\nPHI{01} \!=\! \nPHI{10}$, and these lengths are related
by $\nPHI{00}^2 + \nPHI{01} = 1$. To determine the full geometry of the
probe one therefore only needs to parametrise the intra-space products.

Applying Eqs.(\ref{eq:output_RpPhi}) to the other three products, namely
$\bkPHI{01}{01}$, $\bkPHI{00}{11}$, and $\bkPHI{01}{10}$ (whose calculation
is greatly simplified by the previous orthogonality conditions), one
obtains the desired final constraint,
\begin{equation} \label{eq:d_is_a}
  \bkPHI{01}{10} + \bkPHI{00}{11} = 1 - 2\nPHI{01}^2.
\end{equation}

It follows the probe space can now be parametrised with only two real
parameters, the length $\nPHI{01}$ and one of the two inter-space
products. In order to optimise Eve's measurement, it is handier to
translate these constraints in terms of the vectors $\kpsi{}$. Using
Defs.(\ref{eq:joint_evolution}, \ref{eq:def_of_states},
\ref{eq:joint_evolution_e}), and solving for the $\kpsi{}$'s, one finds
{\small
\begin{subequations}
  \begin{align*}
    \kpsi{uu} &= {
      \cos^2 \!\alpha \kPHI{00} + \sin^2 \!\alpha \kPHI{11} +
      \sin\alpha \cos\alpha ( \kPHI{10} + \kPHI{01} ), } \\
    \kpsi{u\bar{u}} &= {
      \cos^2 \!\alpha \kPHI{01} - \sin^2 \!\alpha \kPHI{10} +
      \sin\alpha \cos\alpha ( \kPHI{11} - \kPHI{00} ), } \\
    \kpsi{\bar{u}u} &= {
      \cos^2 \!\alpha \kPHI{10} - \sin^2 \!\alpha \kPHI{01} +
      \sin\alpha \cos\alpha ( \kPHI{11} - \kPHI{00} ), } \\
    \kpsi{\bar{u}\bar{u}} &= {
      \cos^2 \!\alpha \kPHI{11} + \sin^2 \!\alpha \kPHI{00} -
      \sin\alpha \cos\alpha ( \kPHI{10} + \kPHI{01} ), }
  \end{align*}
\end{subequations}
} and similar relations for signals $v$ and $\bar v$, which, due to the
perfect symmetry of the bases, are not relevant here. Trivial but lengthy
calculations show that the correspondence between the $\kpsi{}$'s and the
$\kPHI{}$'s is unitary (although not so easy to spot, since both vectors
sets are not orthogonal and not normalised), and therefore all vector
products are preserved. 

Since attack optimisation is performed at constant QBER, it is better to
have $e$ as a free variable; this is easily achieved with the following
reasoning. The value of the QBER cannot be changed by a local measurement
by Eve after the signal-probe interaction is terminated, and, by
definition, does not depend on the reconciliation procedure. From
Eq.(\ref{eq:joint_evolution}) it is immediate to understand that, if signal
$\ket{a}$ is sent by Alice, an error shows up at Bob's site with
probability $\bkpsi {a\bar{a}} {a\bar{a}}$. Considering that all signals
have the same a-priori probability of $1/4$, and that the parametrisation,
by construction, satisfies the basis-exchange symmetry, one concludes that
\begin{equation} \label{eq:FPEP_e}
  \newcommand{\myvals}[1]{\makebox[0pt][l]{$\scriptstyle a = #1$}}
  e = \tfrac{1}{4} \sum_{\myvals{u,\bar{u},v,\bar{v}}}
  \bkpsi{a\bar{a}}{a\bar{a}}
  = \tfrac{1}{2} \sum_{\myvals{u,\bar{u}}} \bkpsi{a\bar{a}}{a\bar{a}}
  = \npsi{01}^2.
\end{equation}

Therefore, the vectors of the ``error set'', $\kpsi{01}$ and $\kpsi{10}$
have length equal to $\sqrt{e}$, and the vectors of the ``good set'',
$\kpsi{00}$ and $\kpsi{11}$, have length equal to $\sqrt{1-e}$; moreover,
the inter-space products, $\bkpsi{00}{11}$ and $\bkpsi{01}{10}$, sum up to
$1-2e$. By introducing the inter-space imbalance $\delta$, all these
relations can be summarised as in the following table:
\begin{subequations} 
  \label{eq:values_of_bkpsi_simple} 
  \begin{align}
    \label{eq:cross_products_simple}
    \textrm{Span} \left\{ \kpsi{uu}, \kpsi{\bar{u}\bar{u}} \right\}
    &\perp \textrm{Span} \left\{ \kpsi{u\bar{u}}, \kpsi{\bar{u}u} \right\} \\
    \label{eq:uu2_bubu2_simple}
    \npsi{uu}^2 = \npsi{\bar{u}\bar{u}}^2 &= 1 - e, \\ 
    \label{eq:ubu2_buu2_simple}
    \npsi{u\bar{u}}^2 = \npsi{\bar{u}u}^2 &= e, \\
    \label{eq:uububu_simple} 
    \bkpsi{uu}{\bar{u}\bar{u}} &= \tfrac{1}{2} - e - \delta, \\
    \label{eq:ububuu_simple}
    \bkpsi{u\bar{u}}{\bar{u}u} &= \tfrac{1}{2} - e + \delta. 
  \end{align}
\end{subequations}
The imbalance is also limited by the geometrical constraint of scalar
products, {\em i.e.}, Schwartz inequality. \\[.4ex]
\begin{minipage}{.4\linewidth}
  \includegraphics[width=\linewidth]{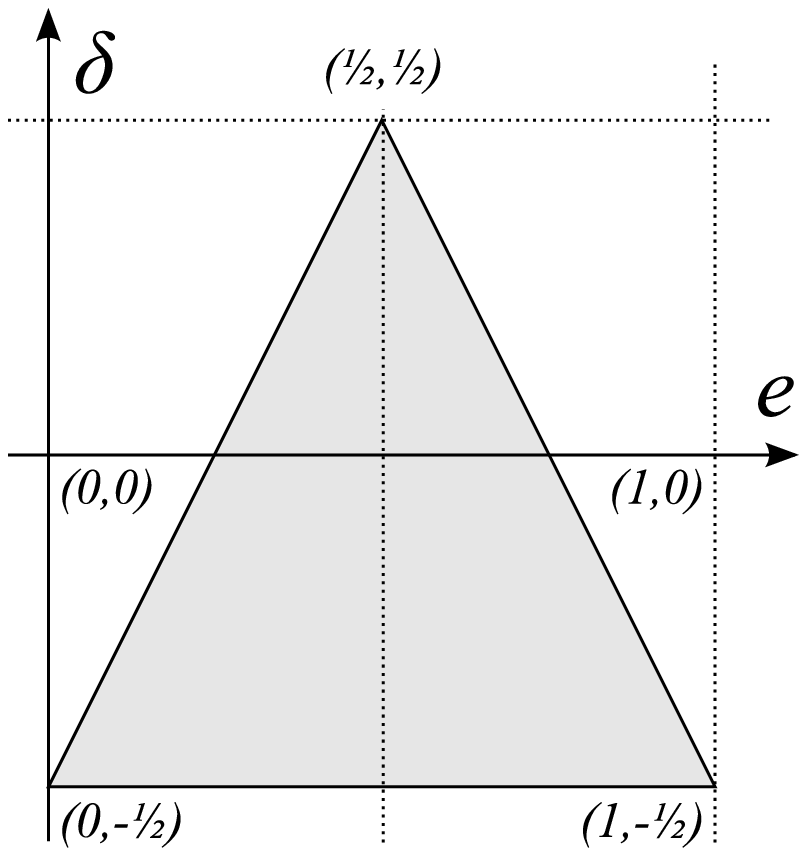}
\end{minipage}
\hfill
\begin{minipage}{.55\linewidth}
  The allowed values for $(e, \delta)$,
  \begin{equation} \label{eq:valid_pairs_ed}
    - \tfrac{1}{2} \leq \delta \leq + \tfrac{1}{2} - | 1 - 2e |,
  \end{equation}
  determined by
  \begin{subequations}
    \begin{align}
      | \tfrac{1}{2} - e - \delta | &\leq 1-e, \\
      | \tfrac{1}{2} - e + \delta | &\leq e,
    \end{align}
  \end{subequations}
  are represented on the left.
\end{minipage}

\smallskip
\noindent In the following of the article the set of equations
(\ref{eq:values_of_bkpsi_simple}) is used, still under the name of FPEP
parametrisation.

\subsection{Estimation of Eve's inference power and the discarded fraction}
\label{sec:discarded_fraction}

As already explained in the introduction, after key reconciliation a
procedure called privacy amplification is applied to reduce Eve's knowledge
to negligible amount (assuming Eve is forced to measure at this
point). Privacy amplification employs universal$_2$ hashing functions to
compress the reconciled key, of length $\bar{n}$, to a final key, of length
$r$. The discarded fraction $\tau$ is then defined as
\begin{equation} \label{eq:def_of_tau}
  \tau = \frac{\bar{n} - r}{\bar{n}}.
\end{equation}
The theory of privacy amplification was developed in a seminal article by
Bennett, Brassard, Cr{\'e}peau and Maurer \cite{BBCM95}, who found a
condition for {\em strong security}. L\"utkenhaus \cite{Lutkenhaus96} used
it to bound Eve's average \footnote{Therefore, QKD security proofs from this
period adopted {\em average strong security} instead of proper {\em strong
security}, as defined in \cite{BBCM95}.} Shannon information on the final
key: for individual attacks, the eavesdropper, on average, knows less than
$1/\ln 2$ bits of the final key provided
\begin{equation} \label{eq:formula_for_tau}
  \tau(e) \geq 1 + \log \cpi,
\end{equation}
where $\cpi$ is the maximum {\em average collision probability} of Eve's
knowledge of {\em one bit} of the reconciled key, for a fixed value of the
disturbance, the QBER $e$. Note that, under conservative assumptions, all
noise on the quantum channel {\em may} be attributed to Eve, but it does
not {\em have to}; therefore, $\tau(e)$ must be a non decreasing
function. If, for instance, $\tau(e'>e) < \tau(e)$, then Eve could perform
the attack causing error $e$, and then pass Bob's signal through a
depolarising channel with error $e'-e$. Therefore, in the following, all
$\tau$'s are to be considered as monotonicised. If $S$ is the random
variable corresponding to the bit sent by Alice, with values $s=0,1$, and
$M$ is the random variable corresponding to all knowledge acquired by Eve,
with values $m$, the $\cpi$ is defined as
\begin{equation}
  \cpi = \sum_{m} P(M=m) \sum_s P^2(S=s|M=m).
\end{equation}

However, when the approach of \cite{SRSF98} is followed, it is not
necessary to calculate the conditional probabilities $P(S=s|M=m)$ nor the
marginal probabilities $P(M=m)$, because the largest possible value of
$\cpi$ can be obtained by direct inspection of the state of Eve's probe
after interaction, as shown in section \ref{sec:Slutsky_approach_with_ED}.

\section{Discarded fraction for individual attacks against a protocol
  using ``faulty bits dumping'' as reconciliation method}
\label{sec:Slutsky_approach_with_ED}

In section \ref{sec:parametrisation} it was shown that the QBER $e$ is
completely determined by the signal-probe interaction during transmission.
This is not the case for Eve's inference power, which depends also on the
reconciliation method. Slutsky \textit{et al.} \cite{SRSF98}, followed by
\cite{Brandt05, KSWS07, ShapiroWong06}, considered only the case when all
errors are discarded from the sifted key.  An evaluation of the cost of
this procedure is postponed to Sec.(\ref{sec:reconciliation}); for the time
being it will be assumed that it can be performed without giving Eve any
piece of information other than the indexes of the retained bits.

Of course, it is very relevant to Eve that reconciliation is performed
through error discard; in fact, her state of knowledge on the signal-probe
system conditioned on Alice sending state $\ket{a}$ changes from that in
Eq.(\ref{eq:joint_evolution}) to a pure state, just as if Bob measurement
had collapsed the signal state into $\ket{a}$,
\begin{equation} \label{eq:collapse}
  U \ket{a} \ww = \ket{a} \kpsi{aa} + \ket{\bar{a}} \kpsi{a\bar{a}}
  \xrightarrow{\textrm{``collapse''}} \ket{a} \kpsi{aa}.
\end{equation}

If, for instance, the encoding basis was $\{ \uo, \ui \}$, Eve's probe, in
Eve's view, would be in an equiprobable mixture of $\kpsi{uu}$ and
$\kpsi{\bar{u}\bar{u}}$. In this case, intuitively, the largest inference
power is given by a measurement that maximises the probability to tell the
first case apart from the second. It is known \cite{Helstrom76, FuchsPhD}
that optimal ambiguous discrimination (corresponding to a minimum of the
probability $\perr$ of making a wrong guess) can be achieved by means of
projective measurements. For two pure and normalised states, $\ket{\phi_0}$
and $\ket{\phi_1}$, assuming, without lack of generality that
$\braket{\phi_0}{\phi_1} \in \mathbb{R}$, and defining the pure-state
fidelity as,
\begin{equation}
  f = | \braket{\phi_0}{\phi_1} |^2 ,
\end{equation}
the optimal von Neumann measurement is defined by the directions
$\ket{\chi_{0/1}} = \theta_{0/1} \ket{\phi_0} + \theta_{1/0} \ket{\phi_1}$,
where
\begin{equation}
  \theta_{0/1} = \frac{\sqrt{1-\sqrt{f}} 
    \pm \sqrt{1+\sqrt{f}}}{2\sqrt{1-f}},
\end{equation}
and the minimum error probability turns out to be $\perr = \frac{1}{2} [1 -
\sqrt{1-f} \,]$. Building on a result of Levitin \cite{Levitin81,
Levitin95}, the authors of \cite{SRSF98} showed\footnote{Actually, the
authors of \cite{SRSF98} where interested only in Shannon and \reni
entropy; the result for the collision probability is implicit in the
inequality for $\cos^2 2\zeta$ at the bottom of the first column of page
2393.} that this measurement also maximises the average collision
probability and the drop in Shannon and \reni entropy, confirming the
intuition. The maximum collision probability turns out to be
\begin{equation} \label{eq:FPEP_best_cpi_via_f}
  \cpi = 1 - \tfrac{1}{2} f.
\end{equation}

Therefore, in the FPEP approach, the problem of optimising Eve's
measurement is really trivial. The optimal attack is that which minimises
the value of $f$ for a fixed value of $e$. Due to the intrinsic basis
symmetry of the method, the value of the fidelity does not depend on the
basis.\footnote{Note that, since the encoding basis is known at measurement
time, Eve can set up different and independent measurements for the two
cases.} Using Eqs.(\ref{eq:uu2_bubu2_simple}) and (\ref{eq:uububu_simple})
one then easily finds
\begin{equation}
  \sqrt{f} = \frac{ | \bkpsi{uu}{\bar{u}\bar{u}} | }
  { \| \psi_{uu} \| \cdot \| \psi_{\bar{u}\bar{u}} \| }
  = \frac{|\tfrac{1}{2} -e -\delta|}{1 - e}.
\end{equation}
which is minimised at fixed $e \leq 1/3$ by $\delta = 2e - 1/2$ [see the
allowed range for $\delta$ in Eq.(\ref{eq:valid_pairs_ed})], yielding:
\begin{equation} \label{eq:FPEP_best_f}
  \min_z\!|_e \sqrt{f} = \frac{1-3e}{1-e} \qquad (e\leq 1/3)
\end{equation}
(if $e > 1/3$, then, with $\delta = 1/2 - e$, the fidelity is exactly zero,
{\em i.e.}, the two cases are perfectly distinguishable). Substituting this
result in Eq.(\ref{eq:FPEP_best_cpi_via_f}), and then into
Eq.(\ref{eq:formula_for_tau}) finally gives the maximum value of the
discarded fraction (implicit in \cite{SRSF98}, and explicitly given in
\cite{ShapiroWong06}),
\begin{align} \label{eq:max_tau_error_discard}
  \tau(e) &= 1 + \log \cpi = \log(2-f) \\
  &= \log \frac{1+2e-7e^2}{(1-e)^2}
  = \log [1 + 4e - 4e^3 + {\cal O}(e^4) ]. \notag
\end{align}
This formula is valid up to $e=1/3$, where the function reaches its maximum
value, $\tau(1/3) = 1$, after which Eve enjoys complete knowledge of the
key established by Alice and Bob (see also the discussion of section
\ref{sec:discarded_fraction}).

\subsection{The Slutsky-Brandt attack}
\label{sec:Slutsky_Brandt_attack}

Kim \textit{et al.} \cite{KSWS07}, following a proposal by Brandt
\cite{Brandt05}, experimentally simulate a particular eavesdropping attack,
the Slutsky-Brandt (SB) attack, that is a specific case of the general FPEP
class previously described. Their practical implementation uses a CNOT gate
as entangling operation, and error-discard as reconciliation procedure.
This attack can be shown to attain the maximum collision probability, as
given by Eq.(\ref{eq:max_tau_error_discard}), and is therefore optimal
within its class.

The SB attack is now shortly recalled. Eve employs a probe system with the
same dimensionality of the signal (a qubit), and the entangling CNOT gate
uses the signal as control and the probe as target. The computational basis
of the CNOT is the same ``symmetric'' basis $\{ \eo, \ei \}$ of
Eqs.(\ref{eq:def_of_states}); with some abuse of notation, the same symbols
$\eo$ and $\ei$ are used to indicate an arbitrary basis in Eve's space.
The initial probe state is
\begin{equation}
  \ww = \tfrac{1}{\sqrt{2}} \Big[ (C+S) \eo + (C-S) \ei \Big],
\end{equation}
where the parameters $S$ and  $C$ are sine and cosine of some angle,
function of the desired QBER $e \leq 1/2$:
\begin{equation}
  S = \sqrt{2 e}, \qquad C = \sqrt{1-2 e}.
\end{equation}

The total system, upon Eve's action, becomes entangled, and its state can
be decomposed according to the definition of Eq.(\ref{eq:joint_evolution}),
giving
\begin{align}
  \kpsi{\substack{uu\\\bar{u}\bar{u}}} &=
  C \, \frac{\eo + \ei}{\sqrt{2}} \pm 
  \frac{1}{\sqrt{2}} \cdot S \, \frac{\eo - \ei}{\sqrt{2}}, \\
  \kpsi{\substack{u\bar{u}\\\bar{u}u}} &= 
  \ket{T_e} \definedas
  \frac{1}{\sqrt{2}} \cdot S \, \frac{\ei - \eo}{\sqrt{2}}. \label{eq:Te}
\end{align}
Similar equations hold in the other basis.  The probability of having an
error is, as expected, $\braket{T_e}{T_e} = S^2/2 = e$. ``Error states'',
that is the states $\kpsi{a\bar{a}}$, are characterised by independence
from the actual signal $a$, as they are always equal to $\ket{T_e}$. As a
consequence of this, when an error takes place, Eve has no information at
all on the transmitted bit -- the entangling unitary is in fact optimised
for protocols which discard errors instead of correcting them.

The inference power of the SB attack can be calculated, as already seen,
from the fidelity of $\kpsi{uu}$ with respect to $\kpsi{\bar{u}\bar{u}}$;
for $e \leq 1/3$, it is identical to that of Eq.(\ref{eq:FPEP_best_f}),
which proves that this attack is optimal in the class of attacks on
protocols which discard errors of the sifted key:
\begin{equation}
  \sqrt{f} = \frac{ | \bkpsi{uu}{\bar{u}\bar{u}} | }
  { \| \psi_{uu} \| \cdot \| \psi_{\bar{u}\bar{u}} \| }
  = \frac{|2C^2 - S^2|}{2C^2 + S^2} = \frac{|1-3e|}{1-e}.
\end{equation}

\section{Reconciliation: error discard versus error correction}
\label{sec:reconciliation}

As emphasised earlier, a QKD protocol, like BB84, can be implemented in
many variants, by adopting different approaches for reconciliation. Each of
these dialects is a protocol on its own, and trivially comparing the
discarded fraction for different protocols makes as much sense as comparing
apples with pears. However, a common benchmark can be found in the length
of the final secret with respect to the length $n$ of the sifted key (not
the length $\bar{n}$ of the reconciled key).

The problem is further complicated by the fact that the
privacy-amplification bound is based on the average collision probability
of the sifted {\em and} reconciled key. If reconciliation is performed in
clear, by exchanging public messages on the classical channel, $\cpi$ of
the sifted key is modified in ways that are very difficult to account
for. For this reason, it is established practice to exchange reconciliation
information in encrypted form, with a one-time pad. This, of course,
requires a previous secret to be shared by Alice and Bob; this secret is
consumed during the execution of the protocol, and must enter the final
balance of secret key production. The alternative approach of exchanging
public messages and then reducing the final key of an equivalent amount
has never been proven to be more efficient, but it is more difficult to
justify theoretically (see, {\em e.g.}, \cite{CachinMaurer97}).

Articles on BB84 with error discard usually do not mention an explicit
procedure for discarding faulty bits; but it is clear that locating {\em
all} errors in the sifted key is exactly as difficult as correcting the
string altogether (since the output of one procedure can be directly used
to implement the other one), which implies a minimum cost $n h(e)$, where
$h$ is the binary entropy function $h(e) = - e \log e - (1-e) \log (1-e)$,
due to the Shannon limit \cite{Shannon48}. The secret gain is therefore at
most
\begin{equation}
  G_d = n(1-e)(1-\tau_d(e)) - nh(e),
\end{equation}
because {\small(1)} the sifted key of length $n$ is reduced to a reconciled
key of length $\bar{n} = n(1-e)$ by discarding the $ne$ errors, {\small(2)}
the reconciled key is compressed by a factor $1-\tau_d$ during privacy
amplification, and {\small(3)} the cost of tight error discard, $nh(e)$,
must be subtracted from the final balance. The subscript ${}_d$ of $\tau$
is meant to remember that this is the discarded fraction in case of
reconciliation through error discard. This gain can be directly compared
with that of protocols with error correction. In the latter case, $\bar{n}
= n$ (no bits are discarded), and $\tau$ becomes $\tau_c$:
\begin{equation}
  G_c = n(1 - \tau_c(e)) - nh(e).
\end{equation}
Obviously, $0 \leq \tau_c \leq \tau_d \leq 1$, because more information is
available to Eve with error discard than with error correction ({\em i.e.},
the location of all bits received as errors, and the fact that all retained
bits were received without errors). One can consider also a case in which
errors are corrected, but the positions of the corrected spots is leaked to
Eve;\footnote{The case of ``leaked errors'' is considered because it
simplifies a lot of calculations, and is anyway an upper bound to the case
of perfectly encrypted error correction.} the previous considerations are
not invalidated. It is immediate to see that error correction is always
better than error discard, because
\begin{equation}
  \frac{G_c - G_d}{n} = (1-e)(\tau_d - \tau_c) + e (1 - \tau_c) \geq 0.
\end{equation}

Therefore, it makes sense to see what happens to the ``optimal BB84
attack'' when reconciliation is done through error correction, a case
analysed in section \ref{sec:Slutsky_Brandt_with_EC}. One may legitimately
think that other reconciliation procedures could lead to an even larger
gain; for instance, an algorithm could select an error-free part of the
sifted string of length $\bar{n}$ by exchanging a message shorter than
$\bar{n} h(e)$, as long as $\bar{n} < n$. The
overall secret gain is most probably not larger than $G_c$, but this
statement has never been formally proved. Other variants might be explored,
like reconciling Alice's key to the sifted key of Bob, instead of the
opposite, or changing both to a third common string, or merging
reconciliation and privacy amplification into a single step, or even
replacing standard privacy amplification with some other procedure in order
to get closer to the $I(A:B)-I(A:E)$ bound. However, one should also
remember that QKD proofs are not after finding the ``optimal'' protocol,
but after proving that a given, probably sub-optimal but reasonably
efficient protocol is secure under some conditions.

Changing the focus from one protocol to another is moreover often not a
good idea because QKD proofs are a lengthy and expensive collective
effort, which must be to some extent restarted when the protocol is
changed. And all this, not to speak of the apparent impossibility to
parametrise the space of ``all possible QKD protocols''. For QKD protocols,
standardisation is more important than optimisation.

\section{The Slutsky-Brandt attack with an error-correction procedure}
\label{sec:Slutsky_Brandt_with_EC}

\begin{figure}[tb]
  \centering\includegraphics[width=.9\linewidth]{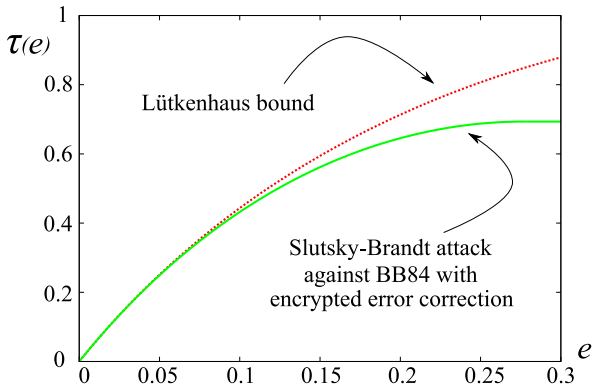}
  \caption{ \label{fig:tau} The fraction of the sifted key that must be
    discarded during privacy amplification in order to counter a SB attack
    against a protocol with encrypted error correction,
    Eq.(\ref{eq:tau_for_SB_with_EC}), versus the QBER $e$ compared with
    L\"utkenhaus bound \cite{Lutkenhaus99}, Eq.(\ref{eq:tau_Lutkenhaus}).
    The first curve reaches its maximum at $e \sim 0.277$, the bound at
    $e=0.5$, where its value is $1$. The curves are non-decreasing, see the
    discussion in section \ref{sec:discarded_fraction}.}
\end{figure}

The SB attack will now be analysed in the context of a BB84 protocol using
encrypted error correction, in order to investigate its claimed optimality.
Because of this choice for reconciliation, the amount of information leaked
to Eve during the raw exchange plus the knowledge of the encoding basis is
all what concerns the calculation of the average collision probability.
Since only individual attacks are allowed, one can consider Eve's activity
as being performed on two separate strings of $n(1-e)$ correct bits and
$ne$ faulty bits respectively. The discarded fraction can thus be written as
\begin{equation} \label{eq:SB_with_EC}
  \tau(e) = (1-e)\tau_{=} + e\tau_{\neq},
\end{equation}
where the first term is related to correct bits and the second one to
faulty bits; this expression is equivalent to
\begin{equation} \label{eq:SB_with_EC_2}
  \tau(e) = 1 + \log \left( \cpiE^{1-e} \cpiD^e \right),
\end{equation}
where $\cpiE$ and $\cpiD$ are the individual average collision
probabilities for error-free and faulty bits respectively. $\tau_{=}$ is
obviously the same quantity determined in section
\ref{sec:Slutsky_approach} for the SB attack, see
Eq.(\ref{eq:max_tau_error_discard}). To calculate the amount of information
leaked to Eve from erroneous bits, note that when the bit measured by Bob
is wrong, the state of the probe collapses to $\ket{T_e}$,
Eq.(\ref{eq:Te}), independently from the bit sent by Alice and the encoding
basis. Therefore, Eve has no mean to distinguish between Alice's two
equiprobable bits, and consequently $\tau_{\neq} = 0$. Using
Eqs.(\ref{eq:SB_with_EC}) and (\ref{eq:max_tau_error_discard}) one finds
\begin{align} 
  \tau(e) &= (1-e) \, \log(1 + 4e - 4e^3 + {\cal O}(e^4)) \notag \\
  &= \log \left(1 + 4e - 4e^2 - 12e^3 + {\cal O}(e^4) \right). 
  \label{eq:tau_for_SB_with_EC}
\end{align}

This discarded fraction can now be compared to the general scenario of
individual attacks considered by L\"utkenhaus in the momentous paper
\cite{Lutkenhaus99}, where the author concludes that $\tau(e)$ is bounded by
\begin{equation} \label{eq:tau_Lutkenhaus}
  \tau(e) \leq \log(1 + 4e - 4e^2).
\end{equation}

In figure \ref{fig:tau}, the discarded fraction necessary to counter a SB
attack is compared with L\"utkenhaus bound (which was not claimed to be
tight). The latter is always higher, hence stronger, than the security
curve derived from the SB attack, the two curves merging only at $e=0$.
For small error rates, most bits are exchanged correctly and, as the SB
attack on correct bits is optimal, the curves converge. When more errors
are introduced, Eve's lack of information on faulty bits weakens her attack.

This shows that in a QKD protocol with encrypted error correction, the SB
attack does not fill the known upper bound, leaving potential room for
stronger individual attacks. The SB curve is however a lower bound, since
the eavesdropping strategy is given explicitly. In the next section, the
question will be investigated whether a stronger FPEP attack can be found,
by appropriately balancing the amount of information Eve can gain from
error-free bits and from bits received incorrectly by Bob.

\section{An optimal attack against BB84 with error correction}

\subsection{With leakage of error positions}
\label{sec:optimal_with_EC_leakage}

This section revisits the FPEP class of attacks against a BB84 QKD protocol
where errors of the sifted key are corrected; however, it is assumed that
the positions of these errors become known to the eavesdropper. This latter
apparently peculiar hypothesis is investigated also in \cite{Lutkenhaus99},
where the author shows that, due to spoiling information, this case can be
used to draw an upper bound also for more secure protocols where Eve has no
information about which bits were received incorrectly by Bob.

The approach to the security proof is very similar to that presented in
section \ref{sec:Slutsky_approach_with_ED}, the difference being, that
here, for a given, known encoding basis, Eve must distinguish between two
pure states for bits received correctly, and two different pure states for
bits received incorrectly, since there are two possibilities for the
``collapse'' of Eq.(\ref{eq:collapse}). For instance, if the basis is $\{
\uo, \ui \}$ and the bit was received incorrectly (which happened with
probability $e$), Eve must distinguish between $\kpsi{u\bar{u}}$ and
$\kpsi{\bar{u}u}$; if the bit was instead received correctly, the two
states are, as before, $\kpsi{uu}$ and $\kpsi{\bar{u}\bar{u}}$. The results
for the second encoding basis are identical, due to the intrinsic symmetry
of the FPEP method. Eq.(\ref{eq:FPEP_best_cpi_via_f}) is thus changed into
\begin{equation} \label{eq:FPEP_best_cpi_via_f_2}
  \cpi = \left( 1 - \tfrac{1}{2} \feq \right)^{1-e}
  \left( 1 - \tfrac{1}{2} \fneq \right)^e,
\end{equation}
with $\feq$ and $\fneq$ defined by the following expressions [which are
then simplified with Eqs.(\ref{eq:uu2_bubu2_simple},
\ref{eq:ubu2_buu2_simple}, \ref{eq:uububu_simple}, \ref{eq:ububuu_simple})],
where the imbalance $\delta$ is constrained by Eq.(\ref{eq:valid_pairs_ed}):
\begin{subequations} \label{eq:def_feq_fdiff}
  \begin{align} 
    \sqrt{\feq} &= \frac{ | \bkpsi{uu}{\bar{u}\bar{u}} | }
	 { \| \psi_{uu} \| \cdot \| \psi_{\bar{u}\bar{u}} \| }
	 = \frac{|\tfrac{1}{2}-e-\delta|}{1-e}, \\
	 \sqrt{\fneq} &= \frac{ | \bkpsi{u\bar{u}}{\bar{u}u} | }
	      { \| \psi_{u\bar{u}} \| \cdot \| \psi_{\bar{u}u} \| }
	      = \frac{|\tfrac{1}{2}-e+\delta|}{e}.
  \end{align}
\end{subequations}

In order to find the optimal attack, it is now sufficient to maximise the
collision probability in Eq.(\ref{eq:FPEP_best_cpi_via_f_2}) over $\delta$.
It is easier to visualise the optimisation problem through the discarded
fraction. In fact, note that
\begin{align}
  \tau &= (1-e) \log(2-\feq) + e \, \log(2-\fneq) \notag \\
  &\leq \log\left[ (1-e)(2-\feq) + e(2-\fneq) \right].
  \label{eq:ineq_tau_leakage}
\end{align}
Finding the maximum $2\delta = -(1-2e)^2$ of the upper bound is trivial
since the argument is a second-degree polynomial in $\delta$. But, for this
value of $\delta$, the two fidelities are equal, and therefore inequality
(\ref{eq:ineq_tau_leakage}) is filled, and the optimisation problem is
solved. One obtains
\begin{gather} 
  \feq = \fneq = f_{\min}(e) = (1-2e)^2, \quad\mathrm{and} \\
  \label{eq:Lutkenhaus_bound}
  \tau(e) = \log\left[ 2-f_{\min}(e) \right] = \log(1+4e-4e^2),
\end{gather}
which is exactly L\"utkenhaus bound of Eq.(\ref{eq:tau_Lutkenhaus}).
Whereas previously this upper bound allowed some margin for lower security
bounds to be found, the present optimisation proves it to be tight when
error positions are leaked. \footnote{From the derivations of section
\ref{sec:parametrisation}, explicit optimal attacks can be devised using
the expressions for the unitary matrix $U$ given by the authors of the FPEP
model \cite{FuchsPeres96}.  Indeed, these authors express the $X$'s in
terms of only four real angles $\{\lambda,\mu,\phi,\theta\}$ in the range
$[0, 2\pi]$. The problem of finding an optimal attack is solved by finding
matrix elements $X$'s, satisfying the conditions Eqs. (\ref{eq:c_is_0}) and
(\ref{eq:d_is_a}), and for which \\[-2ex] \protect\begin{displaymath}
2\delta =-(1-2e)^2 = -2X_0 X_3 - X_1X_2 +2X_5X_6.\protect\end{displaymath}
\\[-3ex] Using the parameters used in \cite{SRSF98}, it follows from
Eqs.(\ref{eq:orthogonality_PHI0010}) and (\ref{eq:c_is_0}) that the
parameters $a,b,c,d$ are further constrained, so that $c=0$, and $a = d =
1-2 e$. The optimisation problem is therefore reduced to the simple task of
finding values of some angles $\{\lambda,\mu,\phi,\theta\}$ for
which\\[-2ex] \protect\begin{displaymath} b = \sin^2\lambda \sin 2\mu +
\cos^2\lambda \sin 2\phi = (1-2e)^2, \protect\end{displaymath} given the
conditions $a=d$ and $c=0$ on \\[-2ex] \protect\begin{displaymath}
\protect\begin{aligned} a &= \sin^2\lambda\sin 2\mu + \cos^2\lambda \cos
2\theta \sin 2\phi, \\ d &= \sin^2\lambda + \cos^2\lambda \cos 2\theta, \\
\textrm{and~~} c &= \cos^2\lambda \sin 2\theta \cos 2\phi.
\protect\end{aligned} \protect\end{displaymath}} Note that, due to the
symmetry $e \leftrightarrow 1-e$, the discarded fraction cannot be a
monotonous curve in this case. Above $e = 50\%$, Eve's tactic for total
knowledge cannot be modelled by the unitary matrix of the FPEP
parametrisation; an additional dissipative evolution on Bob's bit is
necessary.

\subsection{Without leakage of error positions}
\label{sec:optimal_with_EC_no_leakage}

The previous section considered the implementation of a QKD protocol with
error correction and leakage of the positions of the errors, because that
assumption makes the mathematical derivation particularly simple. However,
more secure error-correcting protocols can be devised, in which Eve has no
access to this piece of information. This section investigates whether a
different bound is proper to this instance.

With Eve's assumed lack of knowledge on the error positions, the final
state of the probe after the entangling evolution and the ``collapse'' at
Bob's site is the density matrix $\sigma = \tr_\mathrm{Bob}(\chi)$, with
$\chi$ being the joint state of the probe and the signal. The state
$\sigma$ will be a statistical mixture, over Bob's possible outcomes;
namely
\begin{equation} \label{eq:Eve_denisty_matrix}
  \sigcond{a} = \ketbra{\psi_{aa}}{\psi_{aa}} 
  + \ketbra{\psi_{a\bar{a}}}{\psi_{a\bar{a}}},
\end{equation}
when the input state $\ket{a}$ is sent by Alice; note that $\kpsi{aa}$ and
$\kpsi{a\bar{a}}$ are not normalised; if the normalised vectors were used
instead, the two addends would have a factor $1-e$ and $e$ respectively in
front. Eve must distinguish between the two density matrices ensuing from
the two equiprobable states $\ket{a}$ of Alice, with $a \in \{ u,
\bar{u}\}$.

Suppose that Eve implements the following measurement strategy, on which
there is, a-priori, no claim of optimality. First, she performs a
projective measurement to separate the $\{ \kpsi{uu}, \kpsi{\bar{u}\bar{u}}
\}$ subspace from the $\{ \kpsi{u\bar{u}}, \kpsi{\bar{u}u} \}$ subspace
(finding the first case with probability $1-e$, and the second one with
probability $e$, but this is irrelevant); the separation is possible
because the two subspaces are orthogonal, as shown in
Sec.(\ref{sec:parametrisation}). Then, if the first outcome was found, she
proceeds with the same measurement of
Sec.(\ref{sec:optimal_with_EC_leakage}) for this case, achieving a
collision probability equal to $\feq$; similarly, for the second case, she
achieves $\fneq$. Given that, for these two measurements, both $\feq$ and
$\fneq$ have the same value $f_{\min}(e) = (1-2e)^2$, the average $\cpi$
turns out to be the same as for the case of error correction with leakage
of error positions.

Therefore, there exists a measurement strategy which is ignorant of the
positions of the errors and fills the bound of
Eq.(\ref{eq:Lutkenhaus_bound}). It is however obvious that all attack
strategies that can be implemented without this piece of knowledge can be
implemented also if it is available: in other words, the set of allowed
attacks without leakage is strictly included in the set with leakage, and
therefore, the security bound for the current case cannot exceed the
security bound of Sec.(\ref{sec:optimal_with_EC_leakage}). Thus, the
explicit attack just shown implies that the two bounds are the same, and
that the attack itself is optimal.

It is remarkable that, similarly to Eq.(\ref{eq:FPEP_best_cpi_via_f}), also
in this case the maximum collision probability is linked to the fidelity
\cite{Jozsa94} of the conditional density matrices $\sigcond{u}$ and
$\sigcond{\bar{u}}$. The calculation is greatly simplified by the subspaces
$\{ \kpsi{uu}, \kpsi{\bar{u}\bar{u}} \}$ and $\{ \kpsi{u\bar{u}},
\kpsi{\bar{u}u} \}$ being orthogonal; using Eqs.(\ref{eq:def_feq_fdiff})
one obtains
\begin{multline}
  f(\sigcond{u}, \sigcond{\bar{u}}) = \tr^2 \!
  \sqrt{\!\sqrt{\sigcond{u}} \,\sigcond{\bar{u}} \,\sqrt{\sigcond{u}}}
  = \left[ (1-e)\feq^{\frac{1}{2}} + e\fneq^{\frac{1}{2}} \,\right]^2 \! \\
  = \left( |\tfrac{1}{2} -e -\delta| + |\tfrac{1}{2} -e +\delta| \right)^2
  = (1-2e)^2, \quad
\end{multline}
and therefore $\cpi = 1 - \frac{1}{2}f$. This identity may be true here
only due to the large number of constraints dictated by the symmetries of
the BB84 protocol. However, it would be interesting to know whether the
result holds more generally. This problem is somehow similar to that of
minimum error probability or accessible information. Despite intuition, it
is known \cite{Levitin95, FuchsCaves94} that these two are not equivalent
for mixed states. It is likely that the maximisation of the collision
probability is still a different problem. Formally, the problem would read
like this: provided a flat bit $S$ is transmitted through a quantum
channel, encoded in non-orthogonal density matrices $\rho_0$ and $\rho_1$,
what is the maximum collision probability of the distribution of $S$ that
can be reconstructed by the receiver by means of quantum measurements?

\section{Conclusions}

It has been shown that no real ``threat'' to the security of BB84 QKD
protocols stems from recent developments in implementing an entangling
probe attack. Not only is this attack (claimed to be the ``most powerful
individual attack''\cite{KSWS07, ShapiroWong06}) not threatening the
security bound derived previously by L\"utkenhaus \cite{Lutkenhaus99}, but
it is also shown to be sub-optimal in an efficient and complete QKD
implementation. The SB attack is only an optimal attack for those specific
types of QKD protocols in which the reconciliation procedure is to somehow
discard all faulty bits, which is a less desirable scheme as it leads to a
shorter final shared key.

It should also be pointed out that experiments cannot allow for the
investigation of fundamental security limits, as ``security'' is not an
observable; they can only shed light on the technological feasibility of
specific eavesdropping attacks.

In view of the previous considerations, the recent headline in Nature
purporting that ``quantum cryptography is hacked''
\cite{Brumfiel_NewsNature07, Brumfiel_FeatureNature07} as a result of the
successful implementation of an SB attack is an unfortunate
misunderstanding.\takethisout{and more a concession to sensationalism than
commitment to rigorous scientific haute vulgarisation.} In fact, the
researchers whose work is highlighted in the news feature do not themselves
make any such sensationalistic claim, even though they fail to mention
existing security proofs and do not comment on the consequences their
attack has on existing security bounds.

In this paper it has been shown that improved analysis of FPEP attacks
leads to finding explicit optimal attacks for the case considered in
\cite{Lutkenhaus99}, filling the bound introduced there, which therefore
turns out to be sharp. This holds independently of whether error positions
are leaked to Eve. The analysis gives a simple recipe for devising optimal
individual attacks, the most powerful eavesdropping attacks that could be
implemented with nowadays technology. The complete statement is the
following. An ideal BB84 QKD exchange where the dimensionality of the
signal space is not changed and the imperfection of the experimental
apparatus consists at most in a noisy and lossy channel, and for which
reconciliation through error correction is performed, followed by privacy
amplification, is strongly secure on average against individual attacks if
and only if the discarded fraction $\tau(e)$ satisfies
\begin{displaymath}
  \tau(e) \geq \log(1+4e-4e^2),
\end{displaymath}
(where $e$ is the QBER of the sifted key) both in the case that the
positions of errors come to be known to the eavesdropper, and in the case
that they do not. A byproduct of this analysis is the question whether the
maximum collision probability in distinguishing two mixed density matrices
is always one minus one half of the fidelity of the carrier states.

\section{Acknowledgements}

The authors thanks Andreas Poppe and Anton Zeilinger for interesting discussions
during the preparation of this work. We acknowledge the support of the European Commission through the
integrated projects SECOQC (Contract No. IST-2003-50613) and QAP
(No. 015846) and Austrian Research Centers GmbH - ARC (ITQ
Quantentechnologie). S.~B. is supported by a Hertha-Firnberg fellowship.

\hbadness=10000
\bibliography{optIndivAttack}

\begin{thebibliography}{34}
\expandafter\ifx\csname natexlab\endcsname\relax\def\natexlab#1{#1}\fi
\expandafter\ifx\csname bibnamefont\endcsname\relax
  \def\bibnamefont#1{#1}\fi
\expandafter\ifx\csname bibfnamefont\endcsname\relax
  \def\bibfnamefont#1{#1}\fi
\expandafter\ifx\csname citenamefont\endcsname\relax
  \def\citenamefont#1{#1}\fi
\expandafter\ifx\csname url\endcsname\relax
  \def\url#1{\texttt{#1}}\fi
\expandafter\ifx\csname urlprefix\endcsname\relax\def\urlprefix{URL }\fi
\providecommand{\bibinfo}[2]{#2}
\providecommand{\eprint}[2][]{\url{#2}}

\bibitem[{\citenamefont{Kim et~al.}(2007)\citenamefont{Kim, genannt Wersborg,
  Wong, and Shapiro}}]{KSWS07}
\bibinfo{author}{\bibfnamefont{T.}~\bibnamefont{Kim}},
  \bibinfo{author}{\bibfnamefont{I.~S.} \bibnamefont{genannt Wersborg}},
  \bibinfo{author}{\bibfnamefont{F.~N.~C.} \bibnamefont{Wong}},
  \bibnamefont{and} \bibinfo{author}{\bibfnamefont{J.~H.}
  \bibnamefont{Shapiro}}, \bibinfo{journal}{Phys. Rev. A}
  \textbf{\bibinfo{volume}{75}}, \bibinfo{pages}{042327}
  (\bibinfo{year}{2007}), \eprint{arXiv:quant-ph/0611235}.

\bibitem[{\citenamefont{Slutsky et~al.}(1998)\citenamefont{Slutsky, Rao, Sun,
  and Fainman}}]{SRSF98}
\bibinfo{author}{\bibfnamefont{B.~A.} \bibnamefont{Slutsky}},
  \bibinfo{author}{\bibfnamefont{R.}~\bibnamefont{Rao}},
  \bibinfo{author}{\bibfnamefont{P.-C.} \bibnamefont{Sun}}, \bibnamefont{and}
  \bibinfo{author}{\bibfnamefont{Y.}~\bibnamefont{Fainman}},
  \bibinfo{journal}{Phys. Rev. A} \textbf{\bibinfo{volume}{57}},
  \bibinfo{pages}{2383} (\bibinfo{year}{1998}).

\bibitem[{\citenamefont{Brandt}(2005)}]{Brandt05}
\bibinfo{author}{\bibfnamefont{H.~E.} \bibnamefont{Brandt}},
  \bibinfo{journal}{Phys. Rev. A} \textbf{\bibinfo{volume}{71}},
  \bibinfo{pages}{042312} (\bibinfo{year}{2005}).

\bibitem[{\citenamefont{Shapiro and Wong}(2006)}]{ShapiroWong06}
\bibinfo{author}{\bibfnamefont{J.~H.} \bibnamefont{Shapiro}} \bibnamefont{and}
  \bibinfo{author}{\bibfnamefont{F.~N.~C.} \bibnamefont{Wong}},
  \bibinfo{journal}{Phys. Rev. A} \textbf{\bibinfo{volume}{73}},
  \bibinfo{pages}{012315} (\bibinfo{year}{2006}),
  \eprint{arXiv:quant-ph/0508051}.

\bibitem[{\citenamefont{Brumfiel}(2007{\natexlab{a}})}]{Brumfiel_NewsNature07}
\bibinfo{author}{\bibfnamefont{G.}~\bibnamefont{Brumfiel}},
  \emph{\bibinfo{title}{Quantum cryptography is hacked}},
  \bibinfo{howpublished}{News @ Nature} (\bibinfo{year}{2007}{\natexlab{a}}),
  \bibinfo{note}{online feature (april 27th), whose summary reads: ``{\em
  Simulation proves it's possible to eavesdrop on super-secure encrypted
  messages}''},
  \urlprefix\url{http://www.nature.com/news/2007/070423/full/070423-10.html}.

\bibitem[{\citenamefont{Brumfiel}(2007{\natexlab{b}})}]{Brumfiel_FeatureNature%
07}
\bibinfo{author}{\bibfnamefont{G.}~\bibnamefont{Brumfiel}},
  \bibinfo{journal}{Nature} \textbf{\bibinfo{volume}{447}},
  \bibinfo{pages}{372} (\bibinfo{year}{2007}{\natexlab{b}}), \bibinfo{note}{the
  editor's summary starts with: ``{\em Quantum cryptography is 100\%
  hack-proof. Or at least it was, until the hackers got cracking. Recent
  simulations suggest that it is only a matter of time before a
  quantum-mechanical method of eavesdropping on super-secure encrypted messages
  is developed. \dots}''}.

\bibitem[{\citenamefont{L{\"u}tkenhaus}(1999)}]{Lutkenhaus99}
\bibinfo{author}{\bibfnamefont{N.}~\bibnamefont{L{\"u}tkenhaus}},
  \bibinfo{journal}{Phys. Rev. A} \textbf{\bibinfo{volume}{59}},
  \bibinfo{pages}{3301} (\bibinfo{year}{1999}),
  \eprint{arXiv:quant-ph/9806008v2}.

\bibitem[{\citenamefont{Gisin et~al.}(2002)\citenamefont{Gisin, Ribordy,
  Tittel, and Zbinden}}]{GRTZ02}
\bibinfo{author}{\bibfnamefont{N.}~\bibnamefont{Gisin}},
  \bibinfo{author}{\bibfnamefont{G.}~\bibnamefont{Ribordy}},
  \bibinfo{author}{\bibfnamefont{W.}~\bibnamefont{Tittel}}, \bibnamefont{and}
  \bibinfo{author}{\bibfnamefont{H.}~\bibnamefont{Zbinden}},
  \bibinfo{journal}{Rev. Mod. Phys.} \textbf{\bibinfo{volume}{74}},
  \bibinfo{pages}{145} (\bibinfo{year}{2002}), \eprint{arXiv:quant-ph/0101098}.

\bibitem[{\citenamefont{Du{\v{s}}ek et~al.}(2006)\citenamefont{Du{\v{s}}ek,
  L{\"u}tkenhaus, and Hendrych}}]{DLH06}
\bibinfo{author}{\bibfnamefont{M.}~\bibnamefont{Du{\v{s}}ek}},
  \bibinfo{author}{\bibfnamefont{N.}~\bibnamefont{L{\"u}tkenhaus}},
  \bibnamefont{and} \bibinfo{author}{\bibfnamefont{M.}~\bibnamefont{Hendrych}},
  in \emph{\bibinfo{booktitle}{Quantum Cryptography}}, edited by
  \bibinfo{editor}{\bibfnamefont{E.}~\bibnamefont{Wolf}}
  (\bibinfo{publisher}{Elsevier}, \bibinfo{year}{2006}),
  vol.~\bibinfo{volume}{49} of \emph{\bibinfo{series}{Progress in Optics}},
  chap.~\bibinfo{chapter}{5}, \eprint{arXiv:quant-ph/0601207}.

\bibitem[{\citenamefont{Scarani et~al.}()\citenamefont{Scarani,
  Bechmann-Pasquinucci, Cerf, Du{\v{s}}ek, L{\"u}tkenhaus, and
  Peev}}]{SBPCDL08}
\bibinfo{author}{\bibfnamefont{V.}~\bibnamefont{Scarani}},
  \bibinfo{author}{\bibfnamefont{H.}~\bibnamefont{Bechmann-Pasquinucci}},
  \bibinfo{author}{\bibfnamefont{N.~J.} \bibnamefont{Cerf}},
  \bibinfo{author}{\bibfnamefont{M.}~\bibnamefont{Du{\v{s}}ek}},
  \bibinfo{author}{\bibfnamefont{N.}~\bibnamefont{L{\"u}tkenhaus}},
  \bibnamefont{and} \bibinfo{author}{\bibfnamefont{M.}~\bibnamefont{Peev}},
  \bibinfo{note}{``A Framework for Practical Quantum Cryptography'', in
  preparation}.

\bibitem[{\citenamefont{Bennett and Brassard}(1984)}]{BennettBrassard84}
\bibinfo{author}{\bibfnamefont{C.~H.} \bibnamefont{Bennett}} \bibnamefont{and}
  \bibinfo{author}{\bibfnamefont{G.}~\bibnamefont{Brassard}}, in
  \emph{\bibinfo{booktitle}{Proc. of IEEE International Conference on
  Computers, Systems, and Signal Processing, Bangalore, India}}
  (\bibinfo{year}{1984}), pp. \bibinfo{pages}{175--179},
  \urlprefix\url{http://www.research.ibm.com/people/b/bennetc/bennettc19846979%
0513.pdf}.

\bibitem[{\citenamefont{Ekert}(1991)}]{Ekert91}
\bibinfo{author}{\bibfnamefont{A.~K.} \bibnamefont{Ekert}},
  \bibinfo{journal}{Phys. Rev. Lett.} \textbf{\bibinfo{volume}{67}},
  \bibinfo{pages}{661} (\bibinfo{year}{1991}).

\bibitem[{\citenamefont{Bennett et~al.}(1992)\citenamefont{Bennett, Brassard,
  and Mermin}}]{BBM92}
\bibinfo{author}{\bibfnamefont{C.~H.} \bibnamefont{Bennett}},
  \bibinfo{author}{\bibfnamefont{G.}~\bibnamefont{Brassard}}, \bibnamefont{and}
  \bibinfo{author}{\bibfnamefont{D.~N.} \bibnamefont{Mermin}},
  \bibinfo{journal}{Phys. Rev. Lett.} \textbf{\bibinfo{volume}{68}},
  \bibinfo{pages}{557} (\bibinfo{year}{1992}),
  \urlprefix\url{http://kh.bu.edu/qcl/pdf/bennettc199263611215.pdf}.

\bibitem[{\citenamefont{Maurer}(1993)}]{Maurer93}
\bibinfo{author}{\bibfnamefont{U.~M.} \bibnamefont{Maurer}},
  \bibinfo{journal}{IEEE Trans. Inf. Theory} \textbf{\bibinfo{volume}{39}},
  \bibinfo{pages}{733} (\bibinfo{year}{1993}).

\bibitem[{\citenamefont{Bennett et~al.}(1995)\citenamefont{Bennett, Brassard,
  Cr{\'e}peau, and Maurer}}]{BBCM95}
\bibinfo{author}{\bibfnamefont{C.~H.} \bibnamefont{Bennett}},
  \bibinfo{author}{\bibfnamefont{G.}~\bibnamefont{Brassard}},
  \bibinfo{author}{\bibfnamefont{C.}~\bibnamefont{Cr{\'e}peau}},
  \bibnamefont{and} \bibinfo{author}{\bibfnamefont{U.~M.}
  \bibnamefont{Maurer}}, \bibinfo{journal}{IEEE Trans. Inf. Theory}
  \textbf{\bibinfo{volume}{41}}, \bibinfo{pages}{1915} (\bibinfo{year}{1995}),
  \urlprefix\url{http://puhep1.princeton.edu/
  ~mcdonald/examples/QM/bennett_ieeetit_41_1915_95.pdf}.

\bibitem[{\citenamefont{Inamori et~al.}(2007)\citenamefont{Inamori,
  L{\"u}tkenhaus, and Mayers}}]{ILM01}
\bibinfo{author}{\bibfnamefont{H.}~\bibnamefont{Inamori}},
  \bibinfo{author}{\bibfnamefont{N.}~\bibnamefont{L{\"u}tkenhaus}},
  \bibnamefont{and} \bibinfo{author}{\bibfnamefont{D.}~\bibnamefont{Mayers}},
  \bibinfo{journal}{Eur. Phys. J. D} \textbf{\bibinfo{volume}{41}},
  \bibinfo{pages}{599} (\bibinfo{year}{2007}), \bibinfo{note}{this may be the
  same contribution presented at the NEC workshop on Quantum Cryptography,
  December 1999, without proceedings}, \eprint{arXiv:quant-ph/0107017}.

\bibitem[{\citenamefont{Gottesman et~al.}(2004)\citenamefont{Gottesman, Lo,
  L{\"u}tkenhaus, and Preskill}}]{GLLP04}
\bibinfo{author}{\bibfnamefont{D.}~\bibnamefont{Gottesman}},
  \bibinfo{author}{\bibfnamefont{H.-K.} \bibnamefont{Lo}},
  \bibinfo{author}{\bibfnamefont{N.}~\bibnamefont{L{\"u}tkenhaus}},
  \bibnamefont{and} \bibinfo{author}{\bibfnamefont{J.}~\bibnamefont{Preskill}},
  \bibinfo{journal}{Quant. Inf. Comput.} \textbf{\bibinfo{volume}{4}},
  \bibinfo{pages}{325} (\bibinfo{year}{2004}), \eprint{arXiv:quant-ph/0212066}.

\bibitem[{\citenamefont{Shannon}(1948)}]{Shannon48}
\bibinfo{author}{\bibfnamefont{C.~E.} \bibnamefont{Shannon}},
  \bibinfo{journal}{Bell Syst. Tech. J.} \textbf{\bibinfo{volume}{27}},
  \bibinfo{pages}{379} (\bibinfo{year}{1948}), \bibinfo{note}{this article was
  published in two parts (July and October issues)},
  \urlprefix\url{http://cm.bell-labs.com/cm/ms/what/shannonday/paper.html}.

\bibitem[{\citenamefont{Fuchs and Peres}(1996)}]{FuchsPeres96}
\bibinfo{author}{\bibfnamefont{C.~A.} \bibnamefont{Fuchs}} \bibnamefont{and}
  \bibinfo{author}{\bibfnamefont{A.}~\bibnamefont{Peres}},
  \bibinfo{journal}{Phys. Rev. A} \textbf{\bibinfo{volume}{53}},
  \bibinfo{pages}{2038} (\bibinfo{year}{1996}).

\bibitem[{\citenamefont{Fuchs et~al.}(1997)\citenamefont{Fuchs, Gisin,
  Griffiths, Niu, and Peres}}]{FGGNP97}
\bibinfo{author}{\bibfnamefont{C.~A.} \bibnamefont{Fuchs}},
  \bibinfo{author}{\bibfnamefont{N.}~\bibnamefont{Gisin}},
  \bibinfo{author}{\bibfnamefont{R.~B.} \bibnamefont{Griffiths}},
  \bibinfo{author}{\bibfnamefont{C.-S.} \bibnamefont{Niu}}, \bibnamefont{and}
  \bibinfo{author}{\bibfnamefont{A.}~\bibnamefont{Peres}},
  \bibinfo{journal}{Phys. Rev. A} \textbf{\bibinfo{volume}{56}},
  \bibinfo{pages}{1163} (\bibinfo{year}{1997}),
  \eprint{arXiv:quant-ph/9701039}.

\bibitem[{\citenamefont{L{\"u}tkenhaus}(1996{\natexlab{a}})}]{LutkenhausPhD96}
\bibinfo{author}{\bibfnamefont{N.}~\bibnamefont{L{\"u}tkenhaus}}, Ph.D. thesis,
  \bibinfo{school}{department of Physics and Applied Physics, university of
  Strathclyde, Glasgow} (\bibinfo{year}{1996}{\natexlab{a}}).

\bibitem[{\citenamefont{Waks et~al.}(2002)\citenamefont{Waks, Zeevi, and
  Yamamoto}}]{WZY02}
\bibinfo{author}{\bibfnamefont{E.}~\bibnamefont{Waks}},
  \bibinfo{author}{\bibfnamefont{A.}~\bibnamefont{Zeevi}}, \bibnamefont{and}
  \bibinfo{author}{\bibfnamefont{Y.}~\bibnamefont{Yamamoto}},
  \bibinfo{journal}{Phys. Rev. A} \textbf{\bibinfo{volume}{65}},
  \bibinfo{pages}{052310} (\bibinfo{year}{2002}),
  \eprint{arXiv:quant-ph/0012078}.

\bibitem[{\citenamefont{Stinespring}(1955)}]{Stinespring55}
\bibinfo{author}{\bibfnamefont{W.~F.} \bibnamefont{Stinespring}},
  \bibinfo{journal}{Proc. Amer. Math. Soc.} \textbf{\bibinfo{volume}{6}},
  \bibinfo{pages}{211} (\bibinfo{year}{1955}).

\bibitem[{\citenamefont{Neumark}(1940)}]{Neumark40}
\bibinfo{author}{\bibfnamefont{M.~A.} \bibnamefont{Neumark}},
  \bibinfo{journal}{Izv. Akad. Nauk. SSSR, Ser. Mat.}
  \textbf{\bibinfo{volume}{4}}, \bibinfo{pages}{277} (\bibinfo{year}{1940}).

\bibitem[{\citenamefont{Bru{\ss}}(1998)}]{Bruss98}
\bibinfo{author}{\bibfnamefont{D.}~\bibnamefont{Bru{\ss}}},
  \bibinfo{journal}{Phys. Rev. Lett.} \textbf{\bibinfo{volume}{81}},
  \bibinfo{pages}{3018} (\bibinfo{year}{1998}),
  \eprint{arXiv:quant-ph/9805019}.

\bibitem[{\citenamefont{Landau and Lifshitz}(1981)}]{LandauLifshitzIII}
\bibinfo{author}{\bibfnamefont{L.~D.} \bibnamefont{Landau}} \bibnamefont{and}
  \bibinfo{author}{\bibfnamefont{E.~M.} \bibnamefont{Lifshitz}},
  \emph{\bibinfo{title}{Quantum Mechanics. Non-relativistic Theory}},
  vol.~\bibinfo{volume}{3} of \emph{\bibinfo{series}{Course of Theoretical
  Physics}} (\bibinfo{publisher}{Butterworth Heinemann}, \bibinfo{year}{1981}),
  \bibinfo{edition}{3rd} ed., ISBN \bibinfo{isbn}{978-0-7506-3539-4}.

\bibitem[{\citenamefont{L{\"u}tkenhaus}(1996{\natexlab{b}})}]{Lutkenhaus96}
\bibinfo{author}{\bibfnamefont{N.}~\bibnamefont{L{\"u}tkenhaus}},
  \bibinfo{journal}{Phys. Rev. A} \textbf{\bibinfo{volume}{54}},
  \bibinfo{pages}{97} (\bibinfo{year}{1996}{\natexlab{b}}).

\bibitem[{\citenamefont{Helstrom}(1976)}]{Helstrom76}
\bibinfo{author}{\bibfnamefont{C.~W.} \bibnamefont{Helstrom}},
  \emph{\bibinfo{title}{Quantum Detection and Estimation Theory}}
  (\bibinfo{publisher}{Academic Press}, \bibinfo{year}{1976}), ISBN
  \bibinfo{isbn}{0123400505}.

\bibitem[{\citenamefont{Fuchs}(1996)}]{FuchsPhD}
\bibinfo{author}{\bibfnamefont{C.~A.} \bibnamefont{Fuchs}}, Ph.D. thesis,
  \bibinfo{school}{University of New Mexico} (\bibinfo{year}{1996}),
  \eprint{arXiv:quant-ph/9601020v1}.

\bibitem[{\citenamefont{Levitin}(1981)}]{Levitin81}
\bibinfo{author}{\bibfnamefont{L.~B.} \bibnamefont{Levitin}}, in
  \emph{\bibinfo{booktitle}{IEEE Intern. Symp. on Information Theory}}
  (\bibinfo{address}{Santa Monica, CA, USA}, \bibinfo{year}{1981}),
  \urlprefix\url{http://kh.bu.edu/qcl/pdf/levitinl19740910011f.pdf}.

\bibitem[{\citenamefont{Levitin}(1995)}]{Levitin95}
\bibinfo{author}{\bibfnamefont{L.~B.} \bibnamefont{Levitin}}, in
  \emph{\bibinfo{booktitle}{Quantum Communication and Measurement}}, edited by
  \bibinfo{editor}{\bibfnamefont{V.~P.} \bibnamefont{Belavkin}},
  \bibinfo{editor}{\bibfnamefont{O.}~\bibnamefont{Hirota}}, \bibnamefont{and}
  \bibinfo{editor}{\bibfnamefont{R.~L.} \bibnamefont{Hudson}}
  (\bibinfo{publisher}{Plenum, New York}, \bibinfo{year}{1995}), pp.
  \bibinfo{pages}{439--448}, \bibinfo{note}{proceedings of QCM94}.

\bibitem[{\citenamefont{Cachin and Maurer}(1997)}]{CachinMaurer97}
\bibinfo{author}{\bibfnamefont{C.}~\bibnamefont{Cachin}} \bibnamefont{and}
  \bibinfo{author}{\bibfnamefont{U.~M.} \bibnamefont{Maurer}},
  \bibinfo{journal}{J. Crypt.} \textbf{\bibinfo{volume}{10}},
  \bibinfo{pages}{97} (\bibinfo{year}{1997}),
  \urlprefix\url{http://www.zurich.ibm.com/~cca/papers/link.ps.gz}.

\bibitem[{\citenamefont{Jozsa}(1994)}]{Jozsa94}
\bibinfo{author}{\bibfnamefont{R.}~\bibnamefont{Jozsa}}, \bibinfo{journal}{J.
  Mod. Opt.} \textbf{\bibinfo{volume}{41}}, \bibinfo{pages}{2315}
  (\bibinfo{year}{1994}).

\bibitem[{\citenamefont{Fuchs and Caves}(1994)}]{FuchsCaves94}
\bibinfo{author}{\bibfnamefont{C.~A.} \bibnamefont{Fuchs}} \bibnamefont{and}
  \bibinfo{author}{\bibfnamefont{C.~M.} \bibnamefont{Caves}},
  \bibinfo{journal}{Phys. Rev. Lett.} \textbf{\bibinfo{volume}{73}},
  \bibinfo{pages}{3047} (\bibinfo{year}{1994}).

\end{thebibliography}

\end{document}